# Embedding Ranking-Oriented Recommender System Graphs

Taher Hekmatfar, Saman Haratizadeh, Sama Goliaei
{taher.hekmatfar haratizadeh, sgoliaei}@ut.ac.ir

**ABSTRACT**

Graph-based recommender systems (GRSs) analyze the structural information available in the graphical representation of data to make better recommendations, especially when the direct user-item relation data is sparse. Ranking-oriented GRSs that form a major class of recommendation systems, mostly use the graphical representation of preference (or rank) data for measuring node similarities, from which they can infer a recommendation list using a neighborhood-based mechanism. In this paper, we propose PGRec, a novel graph-based ranking-oriented recommendation framework. PGRec models the preferences of the users over items, by a novel graph structure called *PrefGraph*. This graph is then exploited by an improved graph embedding approach, taking advantage of both factorization and deep learning methods, to extract vectors representing users, items, and preferences. The resulting embeddings are then used for predicting users' unknown pairwise preferences from which the final recommendation lists are inferred.

We have evaluated the performance of the proposed method against the state of the art model-based and neighborhood-based recommendation methods, and our experiments show that PGRec outperforms the baseline algorithms up to %3.2 in terms of NDCG@10 in different MovieLens datasets.

## 1. Introduction

Today with the overload of information, it has been hard to decide about choosing a proper product or service. Recommender systems are appeared to help people in this situation. Collaborative Filtering systems which use user-item interaction has become popular in past years. The neighborhood-based and model-based approaches are two main CF methods. While the neighborhood-based methods try to find similar users to the target user, the model-based ones extract some latent factors of users and items from the user-item interaction data(Shams & Haratizadeh, 2016). Most of the model-based methods apply matrix factorization techniques like NMF(Luo, Zhou, Xia, & Zhu, 2014), PMF(Ma, Yang, Lyu, & King, 2008) and SVD(Sarwar, Karypis, Konstan, & Riedl, 2000) on the user-item interaction matrix. Nevertheless, some researchers have developed new model-based methods with deep learning-based techniques like MLP(L. Zhang, Luo, Zhang, & Wu, 2018), RBM(Fu, Qu, Yi, Lu, & Liu, 2018; Hazrati, Shams, & Haratizadeh, 2019), and AE(W. Zhang, Zhang, Wang, & Chen, 2019). Matrix factorization methods extract the linear relationship between users and items; however, neural network-based methods extract the non-linear relationships which are more effective in prediction or ranking tasks(D. Zhang, Yin, Zhu, & Zhang, 2018). Both methods work fine when there are a lot of user-item interaction records; however, such user-item interactions are sparse in real-world systems(L. Zhang et al., 2018).

One solution to cover the shortcomings of user-item data sparsity is to model the recommender systems' data with graphs and enrich the initial data with some new information extracted from graph structural information(Shams & Haratizadeh, 2016). In recent years different Graph-based Recommender Systems (GRSs) have been suggested. In many neighborhood-based GRSs, the general approach is to extend the concept of neighborhood and similarity among entities by defining and using graph-based distance measures. These methods can directly estimate the proximity of entities to each other by analyzing specially designed paths or random walks in the graph. However, the model-based GRSs, usually need to extract vector representations for entities before they estimate the relations among them. In its simplest form, each entity can be represented by a vector of features, some of which reflect the local structure of the graph around the entity. Nonetheless, more recent researches tend to use deep embedding models to extract better vector representations. In these methods, usually, a preprocessing step is done, for example, using random walks to map the relations among nodes into representations that can be processed by standard embedding techniques such as Skip-Gram. Unfortunately, such a preprocessing step could lose the information, and it is not clear how to do it in a way that the most relevant information from the graph structure is captured in the final vector representations. Fortunately, recent advancements in graph embedding field have made it possible to directly use the graph structure for embedding the entities in a. Such an approach can be especially useful in the GRS domain in which a diverse set of entities and relations are modeled by large heterogeneous graphs.

In this paper, we model the data with a novel heterogeneous graph, which we call PrefGraph. It has fewer nodes in comparison to its predecessors (Shams & Haratizadeh, 2017), and still, it is able to model the implicit and explicit feedback data more precisely. We introduce a graph-based framework for the ranking-oriented recommendation that applies a deep-learning method for direct vectorization of the graph entities as well as predicting the preferences of the users. A large amount of information in the data graph combined with the deep structure of the model provides us the ability to combine the personalization and generalization power in order to achieve high recommendation performance.

We can summarize the main contributions of this paper as follows:
- We introduce PrefGraph, as a new structure for graphical representation of data for the ranking-oriented recommendation.
- We develop a novel method for embedding nodes in a heterogeneous user-item data graph of a GRS. It uses a CNN-based graph embedding technique to fine-tune feature vectors extracted by NMF from a user-item interaction matrix.
- We propose a novel graph-based recommendation framework that uses a deep learning approach for entity embedding and weight prediction. It can be easily applied either if content and side information is available or not, in both implicit and explicit feedback settings.
- Our suggested framework outperforms the state of the art neighborhood-based and model-based baseline algorithms in different benchmark datasets.

The rest of the paper is organized as follows: in section 2, we review some of the techniques in graph-based recommender systems. The main problem in which we are tackling is defined in section 3. Section 4 presents our proposed method to recommend Top-N items in the ranking-oriented recommender system. Finally, the experimental evaluations of our proposed approach with the NDCG metric and some discussions about them are included in section 5.

## 2. Graph-based recommender system

In recent years, graphical models have been widely used in the recommender system domain. In the GRSs (graph-based recommender system), users and items are usually modeled as nodes of the graph, while the interactions between them are represented as edges (Ning, Desrosiers, & Karypis, 2015; Z. Wang, Tan, & Zhang, 2010). Such graphs make it possible to extract new information from direct and indirect relations among nodes that are useful for making better recommendations, especially based on sparse datasets (Tiroshi et al., 2014).

Although some papers have proposed to use homogeneous graphs in the recommender system applications such as recommending new friends in the social network (Silva, Tsang, Cavalcanti, & Tsang, 2010) or identifying similar items in an item-item graph (Joorabloo, Jalili, & Ren, 2019), many kinds of research have used heterogeneous graphs. For example, on

LinkedIn, a heterogeneous graph is the most straightforward method to model the relations among people, companies, groups, educational institutions, job titles, published texts, and comments. Such a heterogeneous graph can then be analyzed in order to recommend users to employers (Vahedian, Burke, & Mobasher, 2017). The simplest form of heterogeneous graphs in recommender systems is a bipartite user-item graph, as shown in Figure 1.

$$\begin{array}{c} & \begin{array}{ccc} i_1 & i_2 & i_3 \end{array} \\ \begin{array}{c} u_1 \\ u_2 \\ u_3 \end{array} & \left[ \begin{array}{ccc} 1 & 0 & 0 \\ 0 & 1 & 1 \\ 1 & 0 & 1 \end{array} \right] \end{array}$$

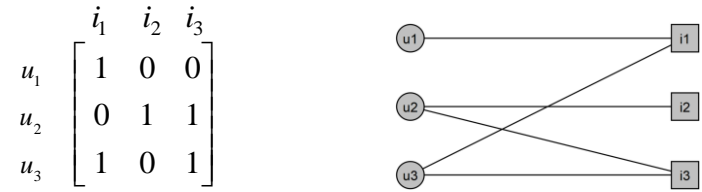

*Figure 1. An example of representing a rating matrix(left) as a bipartite graph(right)*

Even in the simple case of Figure 1, it can be seen that using a graph representation easily extracts useful information that may not be clear in the simple rating matrix: While it is obvious from the matrix that item $i_3$ is a candidate for recommendation to user $u_1$, a graph-based recommender system could also extract the possible interest of user $u_1$ to $i_2$ from the indirect relation between them via $u_1$—$i_1$—$u_3$—$i_3$—$u_2$—$i_2$ path.

Like traditional recommender systems, there are two different approaches in GRSs, too, similarity-based and model-based. Some researches try to calculate the similarity between nodes, and some others try to create a model from nodes and their relationships in the graph. There are different techniques in each approach that we mention some of them here.

Using meta-paths in a heterogeneous graph to define similarity measures is a common approach in many papers. PathSim (Sun, Han, Yan, Yu, & Wu, 2011) which defines the node similarity based on the number of symmetric meth-paths that exist between two nodes and from each node to itself, has been applied in (Yifan Chen, Zhao, Gan, Ren, & Hu, 2016) for Top-N recommendation in the heterogeneous graph. HeteSim (C. Shi, Kong, Huang, Philip, & Wu, 2014) calculates the similarity between two nodes based on the out-neighbors of the first node and in-neighbors of the second node. Unlike the previous two measures, SemRec (C. Shi et al., 2015) considers the weight of edges on defining meta-paths. Random walks and PageRank scores are other measures of similarity in GRSs. One of the first papers that calculates similarities in a heterogeneous user-item-content graph, using random walk is (Fouss, Pirotte, Renders, & Saerens, 2007). They believe that similar nodes will be connected by a larger number of short paths, and long and few paths between nodes demonstrate a huge, difference among them. In (Cooper, Lee, Radzik, & Siantos, 2014) the authors define some scoring method based on random walk in bipartite user-item graph and rank items based on their scores. Their main proposed method is $P^3$, which is the third power of transition matrix $P = D^{-1}A$ in a random walk. In (Yao, He, Huang, Cao, & Zhang, 2015), in addition to content information, a new type of nodes called decision context, which is a combination of time and location, is added to the user-item graph. The weight of the edge between the decision node and other nodes shows the co-occurrence of two nodes. Finally, they apply PPR (Personalized Page Rank) on graph and rank item nodes based on the PPR score. The authors in (Shams & Haratizadeh, 2017) have adopted a tripartite graph structure to model user priorities with a user-item-preference graph and calculates Personalized Page Rank (PPR) to recommend top-n items to each user.

Random walks have also been used for embedding the entities in the recommendation graphs. They optimize the node embeddings in which co-occurring nodes on the short random walks over the graph have similar embeddings (Hamilton, Ying, & Leskovec, 2017). In (Palumbo et al., 2018), Node2Vec (Grover & Leskovec, 2016) as a random walk based embedding technique, generates vectors for user and item nodes. Then the cosine similarity measure provides a ranked list of similar items to each user using the generated vectors. Also, Node2Vec has been used to generate vectors as input for clustering users and items in (J. Chen et al., 2019). Random Walk based approaches usually provide enough flexibility for capturing the local information available in small neighborhoods. Also, they enable the system to explore indirect relations among entities that is useful, especially when dealing with sparse data sets (Goyal & Ferrara, 2018; D. Zhang et al., 2018). However, these methods can be prone to unreliable paths among nodes that can mislead the recommendation process by introducing some noise to the system. So, special care must be taken when using random walks as a base for measuring proximities among nodes in such systems(Shams & Haratizadeh, 2018b).

Matrix factorization(MF) is a traditional embedding method that can be used in GRSs as well. It is possible to apply them on any connections between nodes in the form of node adjacency matrix, Laplacian matrix, node transition probability matrix, or node similarity matrix. HOPE (Ou, Cui, Pei, Zhang, & Zhu, 2016), which uses SVD to learn user-item embedding vectors, is an example of the MF approach. The authors in (Li, Tang, & Chen, 2017) use a tripartite item-user-tag graph. They consider users who tag the same items or use the same tags, as similar and calculate the amount of similarity based on the diffusion. The similarity acts as the regularization term in Regularized Matrix Factorization. MF methods are simple to apply; however, they can be prone to over-generalization and ignorance of fine, local information (D. Zhang et al., 2018).

Also, deep learning methods could embed nodes in the recommender graphs. They can learn the structural information as well as the node information (Fan et al., 2019). Unlike matrix factorization, which is linear, capturing the non-linear relationship between nodes is one of the capabilities of these methods (D. Zhang et al., 2018). Graph Neural Networks (GNNs) are some techniques that could learn representation in any graph including recommender graphs through aggregating information from local neighborhoods using neural networks (Z. Ying et al., 2018). Based on the different types of neural networks, various GNNs are presented (Wu et al., 2020). The Recursive GNNs (RecGNNs) iteratively propagate each node representation (which could be initialized randomly) across the graph, transform received information from neighbor nodes, and aggregate them as a representation for a current node until the node embeddings converge (Scarselli, Gori, Tsoi, Hagenbuchner, & Monfardini, 2008). (Yin, Li, Zhang, & Lu, 2019) and (Fan et al., 2019) are some applications of RecGNNs in recommender systems. Despite the difference in the heterogeneous graph structure, they both propagate information with an attention-based method and aggregate user/item feature information iteratively using an MLP which gets users and items feature vectors as input and provide final vector as output. The Graph Autoencoders (GAEs) could lean node embeddings through an unsupervised manner(Wu et al., 2020). In (Berg, Kipf, & Welling, 2017), the authors have applied an encoder with locality and weight sharing characteristics, on the recommender system bipartite graph which generates the node embeddings and transforms user/item embedding to the first-order neighbor item/user, based on the rating provided by each user for each item. Also, a Bilinear decoder predicts the weights of the user-item edges as unknown ratings. Some researchers have proposed to use the convolution in the aggregation phase of GNN and called it Graph Convolutional Neural Network (GCN) (Bruna, Zaremba, Szlam, & LeCun, 2013; Defferrard, Bresson, & Vandergheynst, 2016; Henaff, Bruna, & LeCun, 2015; Kipf & Welling, 2016). The convolution could define based on the neighborhood in the spatial domain like PinSage algorithm (R. Ying et al., 2018) or it could define in the spectral domain (Kipf & Welling, 2016). The PinSage which is applied to the Pinterest graph, uses a random walk as a sampling method to choose important neighbors and transform their feature vectors through an MLP and aggregate them. In addition to GNNs, some other models apply deep methods on graphs in recommender system domain. Some of them use deep methods to extract representation for user/items like (Hazrati et al., 2019) which some initial feature vectors for users and preferences are extracted from a bipartite preference-based graph then they feed them to an RBM to extract final representation for them or like (Kherad & Bidgoly, 2020) which applies AE to extract initial representation for user/items. Also, there are some researches which embed user/item side information in the knowledge graph like DKN (Deep Knowledge-Aware Network) which applies CNN to embed sentences in each news(H. Wang, Zhang, Xie, & Guo, 2018) or CKE (Collaborative Knowledge Base Embedding) which incorporates SDAE to embed entities textual information(F. Zhang, Yuan, Lian, Xie, & Ma, 2016).

## 3. PGRec

In this section, the proposed method, PGRec (Preference Graph based Recommendation), is presented in detail. The main idea of PGRec is to model a recommender system problem as a weight prediction problem in a new type of heterogeneous graph. It applies this graph structure to extract better representations for nodes using which it predicts some unknown edge weights and infer Top-N recommendations. Figure 2 shows the structure of the method.

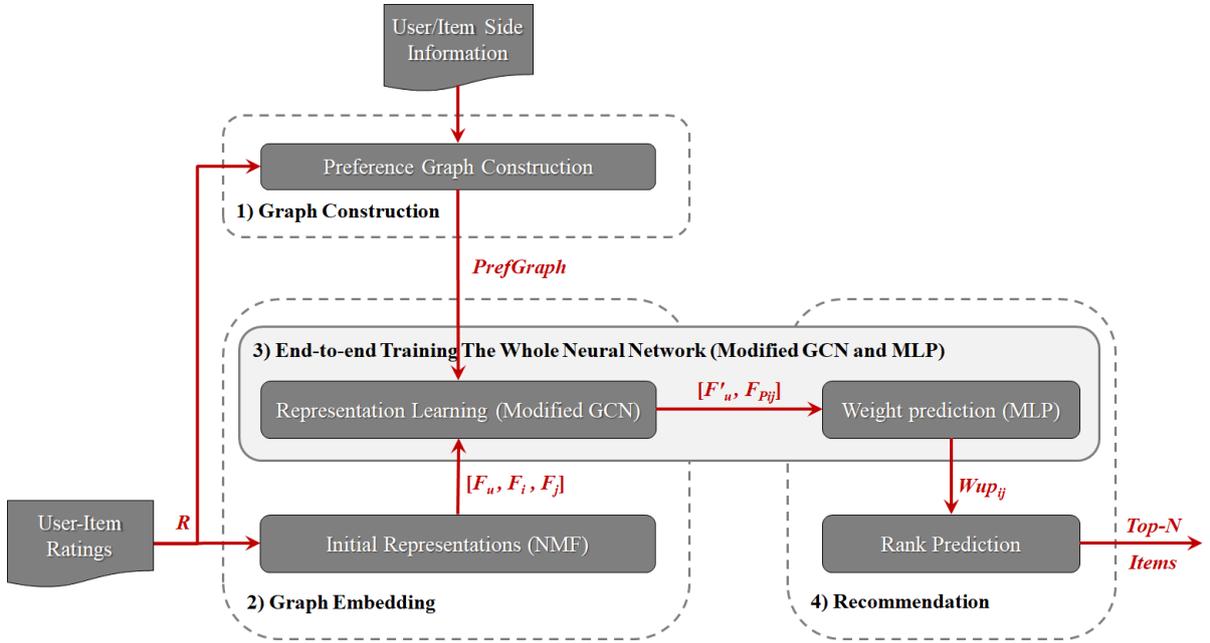

*Figure 2. The proposed method*

In the rest of this section, we first define the problem and then we introduce PrefGraph, the graph structure used to model the data. Then the details of the representation learning method for user and preference nodes in PrefGraph are presented. Finally, the preference prediction and the Top-N recommendation steps are explained.

### 3.1. Problem Definition

Consider the sets of users and items, $U$ and $I$, and suppose that each user $u$ shows his interest to each item $i$ with a rating $r_{ui}$ from $S$ (e.g., $K=[1,5]$ or $K=\{0$ as dislike/no feedback, 1 as like$\}$). There is only one rating for any user $u$ and an item $i$. All these notations are presented in Table 1.

*Table 1. All notations used in the problem and proposed solution*

| Symbol | Descriptions |
|---|---|
| $U$ | Set of all users |
| $I$ | Set of all items |

| Symbol | Descriptions |
|---|---|
| $U_i$ | The subset of users that have rated an item $i$ |
| $I_u$ | The subset of items that have been rated by a user $u$ |
| $N_x$ | The neighbor nodes for target node $x$ |
| $S_x$ | The similar nodes for target node $x$ |
| $r_{min}$ | Minimum feedback possible |
| $r_{max}$ | Maximum feedback possible |
| $R$ | The matrix of user feedbacks to items |
| $r_{ui}$ | Rating or feedback of user $u$ to item $i$ |
| $P$ | The set of pairwise preference |
| $P_u$ | The subset of pairwise preference that user $u$ provides feedback |
| $p_{ij}$ | The pairwise preference of $r_{ui} > r_{uj}$ |
| $f_u$ | The feature vector of user $u$ |
| $f_i$ | The feature vector of item $i$ |
| $v_x$ | The representation vector of node x |
| $G_p$ | The preference graph |
| $G_h$ | The Heterogeneous Graph |

The goal of the recommendation task is to rank the items for a target user and recommend the top ones to him/her. A well-known metric for evaluating the quality of the resulting Top-N recommendation is called Normalized Discounted Cumulative Gain(NDCG)(Järvelin & Kekäläinen, 2000) which is defined as follows:

$$NDCG_u@N = \frac{DCG_u@N}{IDCG_u@N} \quad (1)$$

Where $DCG_u@N$ $DCG_u@N$ is defined as

$$DCG_u@N = \sum_{i=1}^{N} \frac{2^{r_{ui}} - 1}{\log(i+1)} \quad (2)$$

$NDCG_u@N$ shows how good the predicted list is compared to the actual list. $DCG_u@N$ is the DCG value for the Top-N predicted items in descending order and $IDCG_u@N$ is the ideal recommendation score (based on ground truth data). Using IDCG guarantees score 1 for the perfect ranking recommendation.

### 3.2. Graph Construction

We present our method with the help of the data in Figure 3 as a sample of a movie rating dataset, and that contains some information about the users and items of the system.

$$\begin{array}{c} & m_1 & m_2 & m_3 & m_4 \\ u_1 & \begin{bmatrix} 3 & 4 & 5 & ? \\ u_2 & 3 & 5 & ? & 2 \\ u_3 & 4 & 5 & 3 & 3 \end{bmatrix} \end{array}$$

*Figure 3. A user-item interaction matrix*

*Table 2. A subset of user side information*

| Users | Movies |
|---|---|
| $u_1$:(17, Male) | $m_1$:(Fantasy-Drama, 1998) |
| $u_2$:(21, Male) | $m_2$:(Fantasy-Drama, 1998) |
| $u_3$:(20, female) | $m_3$:(Drama-Crime, 1990) |

$m_4$:(Drama-Crime, 1994)

This data could be modeled as a heterogeneous graph like the one in Figure 4.

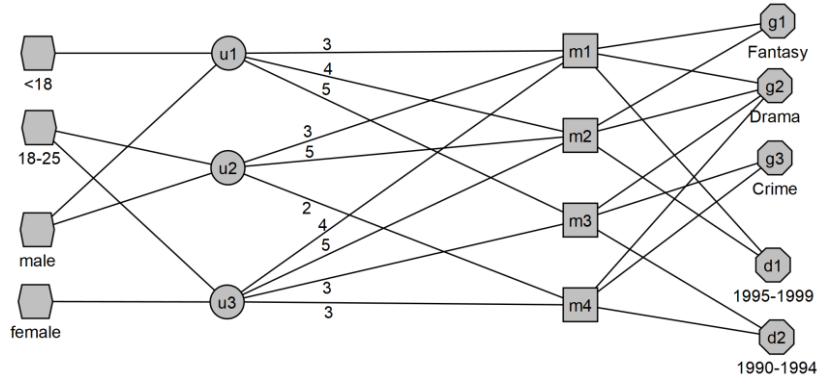

*Figure 4. A heterogeneous graph for recommender system*

It is similar to the bipartite graph in Figure 1 with some content nodes. However, we add a new preference node type to that graph and call it *PrefGraph*. An example of such a graph for corresponding data is presented in Figure 5 which the new node and edge types are distinguished by bold color.

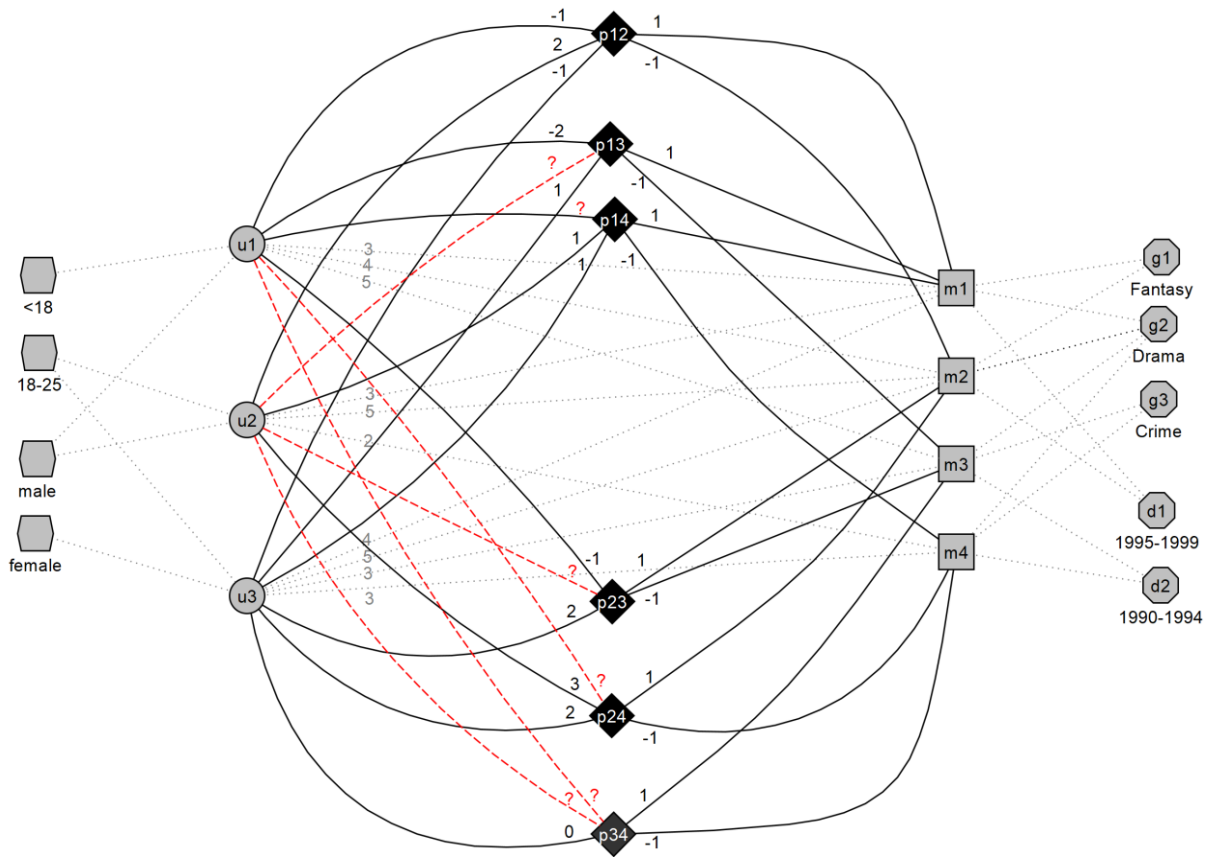

*Figure 5. A sample heterogeneous PrefGraph*

The *PrefGraph*(V, E) is a heterogeneous weighted graph which in its simplest form, is a tripartite graph, in which V = U ∪ I ∪ P and E = $E_{UI}$ ∪ $E_{PI}$ ∪ $E_{UP}$. $E_{UI}$ is the set of weighted edges between users and items based on rating matrix, $E_{PI}$ represents weighted edges connecting a preference node to an item node with weight $W_{PI}$∈{−1,1} ($w_{p_{ij}i} = 1$ and $w_{p_{ij}j} = $

−1). $E_{UP}$ is the set of weighted edges connecting a user node to a preference node with weight $W_{UP} \in [-(k_{max} - k_{min}), (k_{max} - k_{min})]$. $w_{upij} = r_{ui} - r_{uj}$ tells us about how much a user $u$ prefers an item $i$ over another item $j$. Content information about users and items are modeled by adding content nodes. Such nodes are connected with unweighted edges to the corresponding user or item nodes in the graph. In Figure 5, there are some edges with the unknown weights marked as '?'. These edges represent the unknown preferences of users, and we need to predict the weight of such edges, in order to infer the Top-N recommendation lists for users.

### 3.3. Representation Learning

To predict the unknown weights of user-preference edges in Figure 5, we first extract vector representations for each user, item, and preference node. We extract vector representations for users based on information from similar users as well as items rated by the users. Similarly, item representations are generated using similarity information among items, as well as information about users that have rated those items. Finally, the representation for a preference is generated from the vector representations of two items that are connected through that preference. To extract these representations, we propose a hybrid approach based on matrix factorization and deep learning methods. Matrix factorization approaches usually lead to representations that reflect the globally frequent patterns, and combining them with more flexible and sophisticated deep learning methods can hopefully lead to more informative vector representations.

#### 3.3.1. Extracting initial representations

In the recommender system domain, user feedbacks are always non-negative. We have used the Non-Negative Matrix Factorization (NMF) (Luo et al., 2014) to factorize the user-item interaction matrix. This step extracts a set of vector representations for users and items of the system that will later be used by the algorithm as initial representations of the entities. Given an $n \times m$ user-item interaction matrix $R$, NMF will find two non-negative $n \times f$ and $f \times m$ ($f << n,m$) matrices like $F_u$ and $F_i$ in such a way that $R \approx F_u F_i = \tilde{R}$ and each matrix contains f-dimensional vector representations for users or items.

#### 3.3.2. Graph embedding

To consider local relationships between nodes, we apply Convolutional Neural Network (CNN) (Goodfellow, Bengio, & Courville, 2016) on those nodes. Among all deep method algorithms, CNN is known for its attention to local structures of the input. When the input has a graph structure, CNN needs to process the connections among nodes in order to analyze the local spatial correlation among the elements and extract effective local features (Yushi Chen, Jiang, Li, Jia, & Ghamisi, 2016). A typical CNN has a multipart structure with many parameters to learn, and that makes it a powerful tool for extracting complex patterns. Most commonly, CNNs are designed and applied for processing grid structured data like image (L. Zhang et al., 2018); however, graphs usually do not have a grid structure, and every node can generally be adjacent to any other node in the graph. That phenomenon leads to some complications in the process of applying CNNs on graph structures. Recently some researches (Bruna et al., 2013; Defferrard et al., 2016; Henaff et al., 2015; Kipf & Welling, 2016; Niepert, Ahmed, & Kutzkov, 2016) have proposed to use the spectrum of the graph to generalize the concept of convolutional neural networks for the graphical inputs. The suggested framework is called GCN (Graph Convolutional neural networks).

Considering $\mathscr{L}$ as Laplacian matrix of an undirected graph $G$, it is possible to factorize it as $\mathscr{L} = V\Lambda V^T$ where $\Lambda$ is a diagonal matrix of its eigenvalues ($\Lambda_{ii} = \lambda_i$) and the set $V = [v_0, v_1, \ldots, v_{n-1}] \in \mathbb{R}^{n \times n}$ is the matrix of all eigenvectors ordered according to their corresponding eigenvalues (Defferrard et al., 2016). The convolution of a filter $g$ on feature vector node $x$ is defined as follows:

$$x *_G g = V(V^T g \odot V^T x) \quad (3)$$

Equation (3) is the main equation that is used in GCN for implementing the filter convolution over a neighborhood in a graph, while the task of designing good filters remains an active research line. For example, in (Bruna et al., 2013), the filter is replaced with a diagonal matrix, which should be learned, and in (Henaff et al., 2015) the authors considered the filter as smoothing kernel such as splines. In (Defferrard et al., 2016) the authors proposed to apply a polynomial parametric filter $g_\theta(\mathcal{L})$ which $\mathcal{L}$ is Laplacian matrix of an undirected graph $G$ and in (Kipf & Welling, 2016) a first-order approximation of the filter proposed in (Defferrard et al., 2016) has been presented which is the filter we intend to use in our approach. The suggested filter leads to the following convolution operation:

$$Y = X *_G g_\theta = \widetilde{D}^{-0.5} \widetilde{A} \widetilde{D}^{-0.5} X \theta \quad (4)$$

In which $Y \in \mathbb{R}^{n \times f}$ is the convolved signal matrix of the filter matrix $\theta \in \mathbb{R}^{c \times f}$ on a signal matrix $X \in \mathbb{R}^{n \times c}$, $\tilde{A} = A + I_n$ is the normalized adjacency matrix and $\widetilde{D}_{ii} = \sum_j \tilde{A}_{ij}$. This operation could act in the neural network as:

$$H^{(l+1)} = \sigma\big(\widetilde{D}^{-0.5} \tilde{A} \widetilde{D}^{-0.5} H^{(l)} \theta^{(l)}\big) \quad (5)$$

Where H($l$) is the output value in the $l^{th}$ layer (for the first hidden layer $H^{(0)} = X$) and σ(·) denotes a nonlinear activation function.

### 3.3.2.1. Modifying GCN

The proposed convolution operation in equation (5) is simple and effective, however, it has some important drawbacks. The convolution operation in equation (5) for the target node is a kind of taking the average of its neighbor nodes' representations. This operation uses the adjacency matrix or neighbor nodes of the target node. In the adjacency matrix, there is not any difference between neighbor nodes. So, based on this convolution operation all neighbors have the same impact on the final representation of a node. However, in the graph, neighbor nodes are of different types and we may need them to have different impacts on the final representation of the target node. Also, in its current form, the approach defines $\tilde{A} = A + I_n$ which means that the impact of the target node itself on the final representations is the same as the neighbors. However, one can expect that the target node's initial representation has a larger impact on its final representation compared to other nodes.

To handle the aforementioned drawbacks, we propose to adapt the weight matrix $W$ instead of the adjacency matrix. The elements of the weight matrix show the impact of each neighbor node on the target node's representation. Also, we consider weighted self-loops in our graphs as $\widetilde{W} = W + \beta I$, in which $\beta$ is the impact weight of the target node on its final representation. We expect that $\beta$ should be larger than other weights in the weight matrix ($\beta > w_{ij} \in W$). Also, we redefine the degree matrix $\widetilde{D}$ as $\widetilde{D}_{ii} = \sum_{ij} \widetilde{W}_{ii}$. So the modified convolution operation which we use in the layer $H^{(l+1)}$ of our neural network will be as follows:

$$H^{(l+1)} = \sigma\big(\widetilde{D}^{-0.5} \widetilde{W} \widetilde{D}^{-0.5} H^{(l)} \theta^{(l)}\big) \quad (6)$$

In which σ is a Rectifier function as σ($x$) = $x^+$.

### 3.3.2.2. Extracting Item Representation

As it is presented in Figure 6 (a), we define the final representation for an item $i$ as $F'_i$ = concatenate($F_{iS_i}$, $F_{iU_i}$) which means the representation $F'_i$ is defined based on the similar items to item $i$ and also users who have rated it. To calculate the $F_{iS_i}$, as it will be demonstrated in the user/item similarity sub-section (section 3.3.2.5), we define an item-item graph which its edges show the similarity between items and apply the modified GCN in equation (6) on that graph.

Also, we get $F_{iU_i}$ from applying the modified GCN on the user-item interaction subgraph of PrefGraph.

### 3.3.2.3. Extracting Preference Representation
A preference node $p_{ij}$ shows a comparison between two items $i$ and $j$, so we represent a preference node based on vectors of corresponding items as $F_{Pij}$ = concatenate($F'_i$, $F'_j$) as shown in Figure 6 (b). We have considered users and items embedding size as the same. However, because of the concatenation of two item vectors, the size of embedding for preference node is twice the embedding of users and items and in continue, we have to do some operation on both user and preference embedding and their size should be the same. So we pass the concatenated embedding from a single layer neural network to reshape and shrink it half to be the same size as users and items.

### 3.3.2.4. Extracting User Representation
A user $u$ can be presented based on his known preferences. stated previously and also other users who may have similar tastes or characteristics with him. Figure 6 (c) shows the representation of user $u$ as $F_u$ = concatenate($F_{uS_u}$, $F_{uP_u}$). The vector $F_{uS_u}$ is the embedding of similar users to user $u$, resulting from applying the modified GCN in equation (6) on the user-user similarity graph which its construction will be presented in the user/item similarity sub-section (section 3.3.2.5),. The vector $F_{uP_u}$ shows the embedding of preferences that the user $u$ is connected to in PrefGraph. To get this embedding we apply the modified GCN on the user-preference subgraph in PrefGraph.

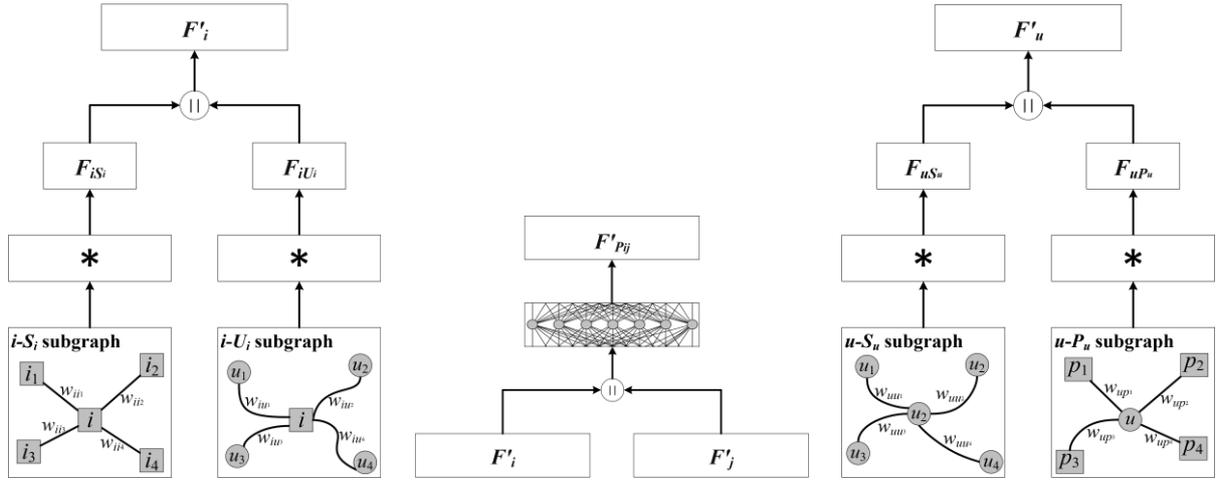

*Figure 6. (Left) Extracting representation for items (Middle) Generating representation for preference nodes (Right) Extracting representation for items. In this figure (\*) shows the modified convolution operation on the input graph in equation (6) and (||) means concatenation of two input vectors.*

### 3.3.2.5. User/Item similarity
As it was mentioned before, part of the representation comes from similar neighbor users/items. By default, there is not any inner edge in users/items layer in PrefGraph, however, it is possible to define some virtual edges and construct user-user/item-item subgraph $G_U(U, E_U, W_U)/G_I(I, E_U, W_I)$. The edges $E_U/E_U$ show similarity between users/items and there are different approaches to define them. Here we consider them as the count of the paths among nodes that are compatible with *User-Content-User* or *Item-Content-Item* meta-paths. The $G_U$ and $G_I$ graphs for the example data in Table 2 is shown in Figure 6:

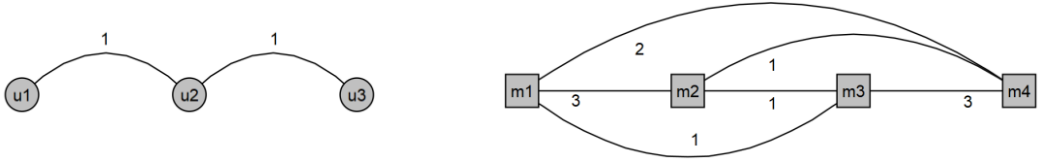

*Figure 7. (Left) The user-user sub-graph. (right) The item-item subgraph*

Because of the limit in the number of contents, these graphs are almost dense, however, all users/items are not completely similar. So, we try to remove some edges in these graphs by clustering nodes. We only keep the edges between nodes that are placed in the same cluster. We apply a recursive two-way spectral clustering algorithm on $G_U$ and $G_I$ which calculates the second eigenvalue $\lambda_2$ of $\mathscr{L}$ and its associated eigenvector $v_2$. Spectral Graph Clustering method considers both feature vectors and the structural information of nodes in the graph. For all node $i \in V_G$, if $v_{2i} > 0$, $i \in cluster_1$, otherwise, $i \in cluster_2$ and it continues by recursively clustering each cluster until $c$ good clusters are found (Nascimento & De Carvalho, 2011). We try to cluster each graph into $c$ well-separated subgraphs which $c$ is different for each dataset and each graph.

It is important to mention that if no content is available, it is possible to use alternative information like co-rating as a similarity measure instead of content-based meta-paths. However, as it will be demonstrated in section 4.7, using content could provide better results in the similarity of nodes.

### 3.4. Preference Prediction

With the representation extracted for user and preference nodes, it is possible to predict the weight between those nodes using a Neural Network with a structure like the one shown in Figure 8.

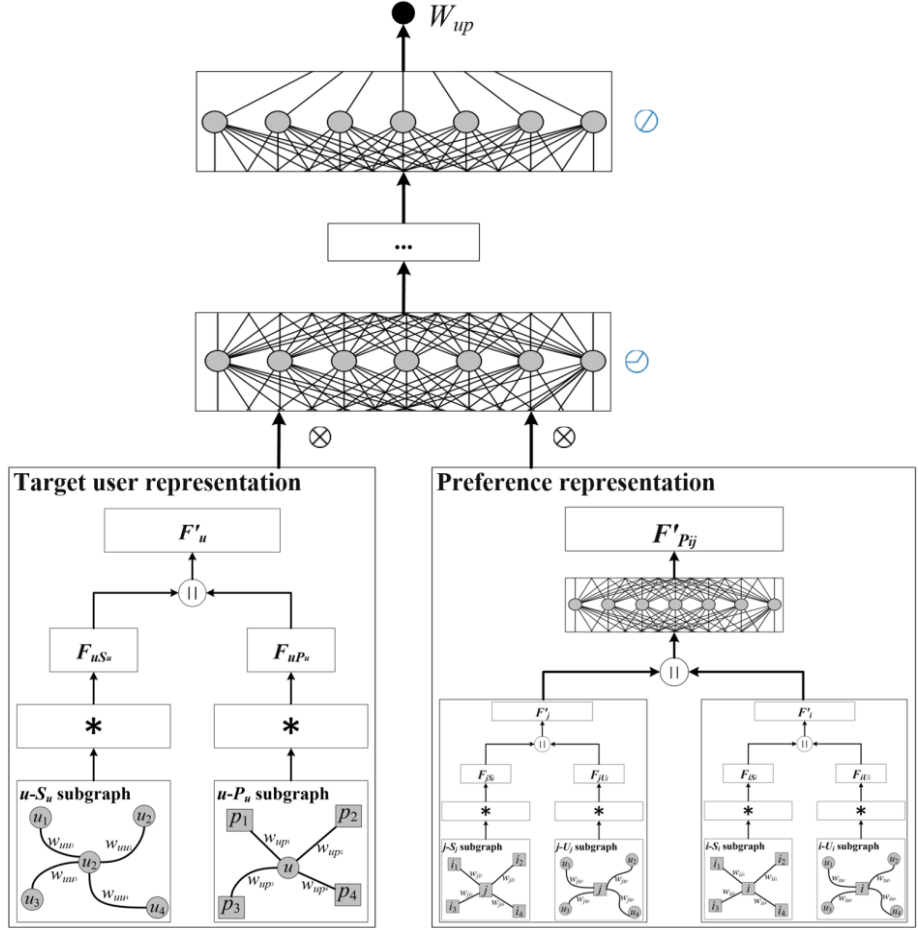

*Figure 8. Predicting weight between user and preference in PrefGraph*

It is a regression network that gets two representations for the target user and the preference node, then concatenates them as a single vector to predict a number as a weight for the edge between the user and preference nodes.

Many deep learning-based GRSs extract final representations first and then apply deep methods on the representations for making predictions. However, an end-to-end training phase may lead to a higher performance according to some studies (Berg et al., 2017). In the end-to-end technique, the embedding and prediction modules are trained together. Following this approach, we train the whole model as a single network and try to minimize the final loss that is the RMSE error of predicted weights for user-preference edges.

### 3.5. Top-N item recommendation

The interest of a user u in an item $i$ can be simply inferred from the predicted weights $w_{up}$ using the following equation:

$$f(u,i) = \frac{1}{(k_{max} - k_{min})(|I_u|-1)} \sum_p w_{up} \times w_{pi} \qquad (7)$$

The $f(u, i)$ shows how much the user $u$ prefers the item $i$ over all other items. It is simply based on multiplying the weight of the user-preference edge and preference-item edge. The maximum value of the $f(u, i)$ is +1, which means the user $u$ prefers the item $i$ over all other items severely, and the minimum value of it is -1, which means the user does not prefer the item $i$ over any other items. Finding Top-$N$ items that maximize $f_u$, ($\arg\max_i f_u$), will provide the $N$ most favorite.

## 4. Experiments and Results

In this section, we explain the experimental setup for the evaluation of our suggested recommendation algorithm against the state of the art methods in the literature.

### 4.1. Datasets

To compare the proposed method with baseline algorithms two public real-world datasets MovieLens100K and MovieLens1M (Harper & Konstan, 2016) which are available at GroupLens website[1] are used. In both datasets, each user has rated at least 20 movies, in a 5-level rating system (1, 2, ..., 5). The details of the datasets are shown in Table 3. In our experiments, we used age, gender, occupation as user side information, and genres and release date as movie side information and ignored other contents like zip code or IMDB URL. To model this data with the graph, we converted the "user age" and "release date" to categorical data by assigning appropriate category names to ages (teenager, young, …) and categorizing the release dates into decades (the 40s, 50s, …).

*Table 3. The details of evaluated datasets*

| Datasets | Users | User side information | Movies | Movie side information | Ratings |
| --- | --- | --- | --- | --- | --- |
| **MovieLens100K** | 943 | Age, gender, zip code, occupation (21 occupations) | 1682 | Title, Movies genres (19 genres), release date, IMDb URL | 99992 |
| **MovieLens1M** | 6041 | Age range, gender, zip code, occupation (21 occupation) | 3779 | Title, Movies genres (18 genres), IMDB URL | 1000162 |

### 4.2. Experimental Setting

We have followed the *weak generalization* methodology as a standard and widely used protocol to evaluate our model. In this method, a fixed number of ratings, called UPL (User Profile Length), is selected randomly from each user feedback set as the training dataset. We have considered UPL={10, 20, 50} in our experiments. The remaining rating data are used as the testing set for evaluate the performance of the model. We removed the users whose UPL size is less than our UPL condition plus 10 as we evaluate the method for predicting the top 10 items per user. So, as an example, for UPL = 20, only users with UPL ⩾ 30 (20 items for the training and 10 items for testing) are considered. Also, the movies that have not appeared in the training set are removed from the test set. The preferences are generated based on training and testing data according to the procedure explained in section 3.2.

### 4.3. Experimental Environment

We trained and tested our model on a PC with the Ubunto 16.04 operating system and Cori5 Intel CPU, 16GB of Ram and an NVIDIA 1050 Ti graphic card with 4GB memory. Because of good machine learning libraries in Python like Pandas and Numpy, we implemented our proposed approach using version 3.5 of this language and Keras library with Tensorflow 1.13. We used version 1.9.7 of Anaconda to simplify package management. Using GPU speeded up the execution of the training and testing phases, more than 20 times.

### 4.4. Evaluation Measure

To evaluate the performance of our developed model, we adopt the NDCG@n metric which $n$ is the number of top-ranked items. On each test run, the average NDCG score of all users is calculated as the performance of the model. For each UPL, we repeated the process of splitting the data into train and test sets, training and performance evaluation 5 times and reported the average performance of the model over those 5 independent experiments for NDCG@5 (recommending 5 best items) and NDCG@10 (recommending 10 best items).

---

[1] https://grouplens.org/datasets/movielens/

### 4.5. Baselines

We have compared the proposed method with some state of the arts different methods in which some are factorization based, some are graph-based and some are neural network-based. It is important to note that our algorithm is designed for making recommendations based on explicit feedback data, for the evaluations to be fair, baseline methods must be selected that can directly handle the explicit feedbacks. So we selected the following set of baseline algorithms that include classic and state of the art methods and ignored some other methods that like NCF (He et al., 2017), NeuRec (S. Zhang et al., 2018), GCF (Yin et al., 2019), and DeepCF (Deng, Huang, Wang, Lai, & Yu, 2019) which can only make a recommendation based on implicit feedback data:

- NMF (Luo et al., 2014) is one of the factorization based algorithms which is described in detail previously.
- PMF (Ma et al., 2008) is a scalable recommender system. It models ratings as a Gaussian distribution. Unlike the most factorization methods which consider unknown values as 0 to reproduce the entire rating matrix, PMF just uses the known values.
- SVD++ (Koren, 2008) is a variant of SVD(Sarwar et al., 2000) which incorporates implicit information and generally acts better than SVD. It is developed by Koren to get Netflix prize. The Netflix dataset is explicit (rating of users to movies), so Koren considered the action of providing a rating as implicit feedback, regardless of its value.
- EigenRank (Liu & Yang, 2008) is one of the first memory-based ranking-oriented CF methods. It considers users' similarities as the Kendall Rank Correlation Coefficient between their rankings of the items and calculates the item ranking-oriented on a random walk on preference information.
- CofiRank (Weimer, Karatzoglou, Le, & Smola, 2008) is one of the first methods which adapt matrix factorization in item ranking task. It tries to maximize NDCG, while, the loss function is replaced with a regularized matrix factorization method.
- SibRank (Shams & Haratizadeh, 2016) is a ranking-oriented recommender system which provides recommendation based on users/items similarity in a preference Bipartite graph.
- GRank (Shams & Haratizadeh, 2017) is a graph-based method that adapts similarity rules on the tripartite graph. Users who have the same opinions about some pairwise comparisons are similar. Items that are similarly favored/disfavored by the same users are similar.
- PreNIT (Shams & Haratizadeh, 2018a) is an item-based collaborative ranking algorithm that models user preferences as labeled edges in the bipartite graph. It recommends items that are similar to the user's preferred items.
- DCR (Hu & Li, 2017) considers the ratings {1, 2, ..., $S$} as $S$ ordinal categorical labels in which each category is presented with a binary matrix. By applying matrix factorization, DCR predicts the probability of each item in each class.
- ListRank-MF (Y. Shi, Larson, & Hanjalic, 2010) is one of the states of the art ranking-oriented collaborative filtering methods that minimizes loss function over matrix factorization.
- BoostMF (Chowdhury, Cai, & Luo, 2015) is a ranking-oriented method that decomposes given $S$-level rating matrix to $S$ binary matrices and applies matrix factorization on them and calculates the probability that a rating predicted as each label and adapt pointwise ranking on them.
- Wide&Deep (Cheng et al., 2016)[2] is a joint double component model. The wide component is a single layer perceptron that can memorize historical data, while the deep component which is an MLP, can generalize the model and predict unseen data.

### 4.6. Parameter Settings

To get the best result of the algorithm, we first trained the algorithm using the *k*-fold technique with random weight initialization. We considered $k = 5$ which means that 20% of training data will be used as a validation set. After pre-training, we used the model parameters as the initialization and then used the whole training set to train the final model. The whole parameter settings are presented in Table 4.

To avoid overfitting, we use three different techniques which are: regularization, dropout, and batch normalization. The regularization penalizes the coefficients of a neural network to reduce the complexity of the model (Wellner et al., 2017). The dropout adds some noise to units in the training phase (Srivastava, Hinton, Krizhevsky, Sutskever, & Salakhutdinov, 2014) and batch normalization standardizes inputs of a unit by changing the distributions of them (Ioffe & Szegedy, 2015).

*Table 4. The parameter settings*

| Parameter | Value |
| --- | --- |
| Learning rate for Adam optimizer | 0.0001 |
| $L^2$ Regularization | 0.0055 |
| Dropout rate | 0.4 in the middle layers to 0.8 in the last layer |
| Embedding sizes | 64 |

### 4.7. Comparison Result

The performance of PGRec and other baseline algorithms in each dataset is presented in Table 5 and Table 6. The recommendation performance of PGRec is reported over 5 independent experiments, but for the baseline algorithm, their best result has been reported. Three different variations of PGRec called PGRecC, PGRecCo, and PGRecS are presented which in these tables. These methods refer to the content-based PGRec, co-rating based PGRec, and the simple form of PGRec without any side information or co-rating relations.

*Table 5. Performance comparison of the proposed method and baseline algorithm in terms of NDCG for the 100K MovieLens dataset.*

| Method | UPL = 10 | | UPL = 20 | | UPL = 50 | |
| --- | --- | --- | --- | --- | --- | --- |
| | **Top5** | **Top10** | **Top5** | **Top10** | **Top5** | **Top10** |
| NMF | 0.605 | 0.635 | 0.611 | 0.618 | 0.621 | 0.613 |
| PMF | 0.633 | 0.661 | 0.676 | 0.686 | 0.676 | 0.682 |
| SVD++ | 0.681 | 0.705 | 0.683 | 0.685 | 0.687 | 0.680 |
| EigenRank | 0.572 | 0.600 | 0.643 | 0.656 | 0.697 | 0.697 |
| CofiRank-NDCG | 0.602 | 0.631 | 0.603 | 0.617 | 0.609 | 0.616 |
| CofiRank-Ordinal | 0.573 | 0.605 | 0.582 | 0.607 | 0.615 | 0.627 |
| SibRank | 0.622 | 0.650 | 0.660 | 0.672 | 0.711 | 0.710 |
| GRank | 0.593 | 0.624 | 0.642 | 0.658 | 0.719 | 0.717 |
| PreNIT | 0.663 | 0.685 | 0.698 | 0.709 | 0.713 | 0.718 |
| DCR | 0.681 | 0.690 | 0.693 | 0.708 | 0.716 | 0.724 |
| ListRank-MF | 0.672 | 0.683 | 0.684 | 0.694 | 0.687 | 0.697 |

---

[2] It is implemented as an official model in TensorFlow repository at
www.github.com/tensorflow/models/tree/master/official/r1/wide_deep

| Method | UPL = 10 | | UPL = 20 | | UPL = 50 | |
|---|---|---|---|---|---|---|
| | Top5 | Top10 | Top5 | Top10 | Top5 | Top10 |
| BoostMF | 0.672 | 0.703 | 0.692 | 0.701 | 0.711 | 0.713 |
| Wide&Deep | 0.646 | 0.677 | 0.658 | 0.663 | 0.659 | 0.662 |
| PGRecS | 0.673 | 0.701 | 0.696 | 0.709 | 0.710 | 0.721 |
| PGRecCo | 0.686 | 0.709 | 0.719 | 0.724 | 0.727 | 0.739 |
| PGRecC | **0.699** | **0.731** | **0.723** | **0.733** | **0.741** | **0.748** |

*Table 6. Performance comparison of the proposed method and baseline algorithm in terms of NDCG for 1M MovieLens dataset.*

| Method | UPL = 10 | | UPL = 20 | | UPL = 50 | |
|---|---|---|---|---|---|---|
| | Top5 | Top10 | Top5 | Top10 | Top5 | Top10 |
| NMF | 0.648 | 0.659 | 0.656 | 0.661 | 0.674 | 0.665 |
| PMF | 0.681 | 0.684 | 0.703 | 0.704 | 0.722 | 0.718 |
| SVD++ | 0.712 | 0.727 | 0.737 | 0.741 | 0.746 | 0.748 |
| EigenRank | 0.606 | 0.614 | 0.700 | 0.699 | 0.693 | 0.692 |
| CofiRank-NDCG | 0.685 | 0.685 | 0.676 | 0.685 | 0.642 | 0.645 |
| CofiRank-Ordinal | 0.630 | 0.637 | 0.643 | 0.646 | 0.670 | 0.675 |
| SibRank | 0.669 | 0.674 | 0.701 | 0.701 | 0.727 | 0.723 |
| GRank | 0.640 | 0.655 | 0.692 | 0.694 | **0.757** | 0.755 |
| PreNIT | 0.713 | 0.718 | 0.735 | 0.736 | 0.741 | 0.738 |
| DCR | 0.719 | 0.726 | 0.744 | 0.743 | 0.754 | 0.762 |
| ListRank-MF | 0.701 | 0.714 | 0.720 | 0.729 | 0.728 | 0.733 |
| BoostMF | **0.743** | 0.743 | 0.738 | 0.748 | 0.752 | 0.751 |
| Wide&Deep | 0.735 | 0.739 | 0.736 | 0.730 | 0.740 | 0.732 |
| PGRecS | 0.726 | 0.739 | 0.740 | 0.744 | 0.741 | 0.750 |
| PGRecCo | 0.731 | 0.748 | **0.750** | 0.757 | 0.749 | 0.755 |
| PGRecC | 0.737 | **0.752** | **0.750** | **0.764** | 0.753 | **0.768** |

The following facts can be observed in the results:
- The PGRecC has achieved the highest NDCG scores on almost every experiment on both MovieLens 100K and 1M datasets. However, the NDCG scores for PGRecS and PGRecCo are lower than some baseline algorithms in both datasets.
- Content-based PGRec performs better than co-rating-based PGRec and the simple form of PGRec in all experiments. Its performance in MovieLens 100K is 3.7% and 1.6% higher than the other two algorithms. Also, it improves NDCG scores of the other two algorithms by 1.8% and 0.7% in MovieLens 1M.
- In MovieLens 100K, PGRecC beats all baseline methods. In MovieLens 1M, PGRecC acts better than other algorithms except the BoostMF in predicting top5 items in UPLs 10 and the GRank in predicting top5 items in UPLs 50.
- In MovieLens 100K, PGRecC on average improves GRank and PreNIT which are baseline GRS algorithms by 11% and 4.5%, Wide&Deep as deep learning-based method by 10.4%, SVD++, DCR and BoostMF as baseline factorization-based methods by 6.2%, 3.9%, and 4.4%. Also, in MovieLens 1M, PGRecC on average acts better than GRank and PreNIT which are baseline GRSs by 8.3% and 3.3%, Wide&Deep as a deep learning-based method by 2.6%, SVD++, DCR and BoostMF as baseline factorization-based methods by 2.6%, 1.8%, and 1.1%.

### 4.8. Discussion

We have provided three different variations of PGRec which in the simplest form there is not any intra-layer relation between nodes in the users'/items layer, so users are just represented based on preferences and items are represented by users only. However, in the two others, we consider some relationships between users/items based on co-rating or side information, and the difference in the NDCG result of them shows the impact of enriching the basic user-item information. However, it has some computational cost and the simple form of PGRec could be trained faster than the PGRecC and PGRecCo by 19% and 23% in UPL=50 on MovieLens 100K. So it is a trade-off between computational cost and accuracy.

Between those three variations, the best NDCG scores belong to PGRecC which is faster than PGRecCo too, because calculating the co-rating relationship has more computational cost than using content information. So, in this section, we discuss PGRecC. First, we talk about the performance of PGRecC. Like most of the model-based methods, in the sparser datasets (lower UPLs) (Aggarwal, 2016), PGRec improves NDCG significantly over neighborhood-based methods like GRank and SibRank. However, the difference between these methods reduces as training datasets grow in larger UPLs, and GRank supers PGRec in Top@5 item recommendation of UPL = 10 in MovieLens 1M.

The PGRecC takes 505 seconds (1.4 seconds for each user) of training in UPL=50 on MovieLens 100K. It takes 8.6 seconds to recommend Top@10 items to each user. It is slower than some simple and very fast model-based algorithms like SVD++ which could create the model in less than 1 second and recommend Top@10 items for each user in less than 10 milliseconds. However, its training time is acceptable in comparison with ranking-oriented model-based algorithms like CofiRank-NDCG which takes 199 seconds to create the model and 0.1 seconds to recommend Top@10 items to each user. One reason for the slower recommendation phase of PGRec is that it needs to predict weights for user-preference edges before it can infer the recommendation list. Since the number of preferences is quadratic in the number of items, PGRec has to make many more predictions compared to algorithms that directly predict user-item relations.

In comparison to other GRSs, PGRecC recommendation time is better than some like SibRank and worse than some others like GRank which both use preference data for item prediction. SibRank and GRank need 32.1 and 2.1 seconds to recommend items to each user. In preference-based GRSs, the most important factor that affects the computational time is the structure of the graph. Small and simpler graphs could reduce computational time.

To verify the effect of embedding size on the result of our experiments, we selected various values in the power of 2 for initial embeddings of users and items. Embeddings with the size of power of 2 are more efficient in time and space. The results as the RMSE loss function in predicting the weights of user-preference edges in UPL = 50 in 100K MovieLens dataset are presented in Figure 9.

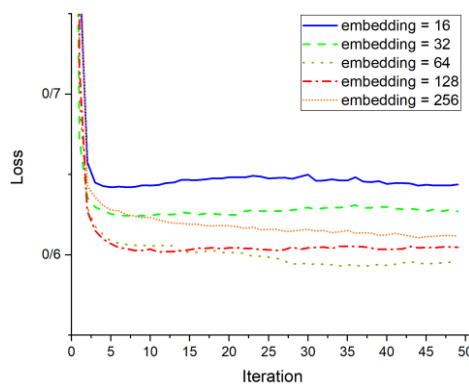

*Figure 9. Train loss in each epoch for different initial embedding size in UPL = 50 of 100k MovieLens dataset*

The result of experiments shows that embedding size 16 and 32 are too small to contain enough information for predicting train dataset. However, increasing embedding size to 64 could improve the performance of the algorithm, while embedding over 128 causes the algorithm to lose performance. The performance of 64 and 128 is almost similar, so it makes more sense to use smaller embedding which is computationally cheaper. Very large embeddings usually lead to overfitting unless a very large train dataset is available. It is worthy to mention that the algorithm starts to converge fast in less than 10 iterations.

Also, we did another experiment to check how NDCG changes on different Top-N recommendation in different UPLs. We considered $N$ as $\{1, 2, …, 10\}$ and UPLs as $\{10, 20, 30, 40, 50\}$. Figure 10 shows the result of this experiment on the 100K MovieLens dataset.

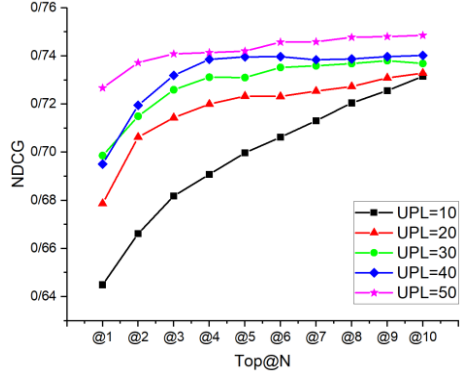

*Figure 10. NDCG improves in larger UPLs and Top-N recommendation*

There are some interesting characteristics of the presented results. First, the NDCG value in Top-1 is in a wider range for different UPLs, however, the range is getting smaller as $N$ increases. UPL = 50 improves NDCG of UPL = 10 more than 13% in Top@1 item recommendation however, this improvement is 6% for Top@5 and 2% for Top@10 item recommendation. It seems that increasing the size of training data does not change the items that are recommended but it can lead to some slight changes in the order of them. So, if it is possible to present a larger list of the recommended items to the user, one can prefer to use smaller samples of data for training to speed up the training phase.

## 5. Conclusion

In our experiments, we developed a model-based GRS for the ranking-oriented recommendation that directly uses a graphical model of data for embedding the entities and predicting the users' unknown pairwise preferences. The proposed graph structure that is used to represent the preference data is smaller than previous ones and is able to model the intensity of preferring one item over another one. The embedding module adopts the modifies GCN, a method to apply deep algorithms on the graph, to represent user, item, and preference nodes. We modified GCN to support a heterogeneous network in which the local information available in a neighborhood, comes from different node types with possibly different degrees of importance. The modified GCN improves the initial NMF-generated representations to achieve more reliable embeddings for entities. The resulted embeddings are used to predict the unknown pairwise preferences and generating the final recommendation list.

The experiments show that the proposed innovations in graphical modeling of data and the embedding and prediction process lead to significant improvements over the state of the art baseline model-based and neighborhood-based recommendation methods. However, since the suggested approach is designed for the ranking-oriented recommendation, its runtime is rather high compared to some baseline algorithms. Applying PGRec as a general GRS framework for the ranking-oriented recommendation in other recommendation domains with more diverse types of entities can be the subject of further research. Also PGRec, like many other methods that infer the recommendation list from predicted pairwise preferences, deals with the problem of high recommendation time and trying to find a solution for that problem is another possible direction for further research.